\documentclass[
reprint,
 amsmath,amssymb,
 aps,
 pre,
]{revtex4-1}

\usepackage{graphicx}
\usepackage{dcolumn}
\usepackage{bm}
\usepackage{hyperref}
\usepackage[mathlines]{lineno}
\usepackage[dvipsnames]{xcolor}
\usepackage{ulem} 

\begin{document}

\preprint{APS/123-QED}

\title{
Statistical properties of the energy flux between two NESS thermostats.}

\author{Mona Lam\`eche}
\author{Antoine Naert}%
\email{Antoine.Naert@ens-lyon.fr}
\affiliation{%
Univ Lyon, ENS de Lyon, Univ Claude Bernard, CNRS, Laboratoire de Physique, F-69342 Lyon, France 
}%

%
%

\date{\today}

\begin{abstract}
We address the question of transport of heat, in out-of-equilibrium systems. The experimental set-up consists in two coupled granular gas Non-Equilibrium Steady State (NESS) heat baths, in which Brownian-like rotors are imbedded. These rotors are electro-mechanically coupled thanks to DC micro-motors, through a resistor $R$, such that energy flows between them. The average flux depends linearly in the difference of the baths' temperature. \\
Varying $R$ allows to extrapolate in the non-dissipative coupling limit ($R\rightarrow0$). We show that, in this limit, the heat flux obeys the Fluctuation Theorem, in a form proposed by Jarzynski and W\'ojcik in $2004$ for the fluctuations of the flux between finite size equilibrium heat baths. 
\end{abstract}

\pacs{Valid PACS appear here}
\maketitle


\section{\label{sec:level1}
Introduction
}
In dissipative systems, Non-Equilibrium Steady States (NESS) result from the balance, in the average, between the work supplied and the heat dissipated, per time unit. (See for instance turbulent flows, granular gases, etc.) In many physical systems of interest, work is injected from the boundaries as dissipation occurs in the bulk. 
The simplest experimental set-up one can think of to investigate the energy transport in NESS systems, is two granular gas thermostats, at distinct effective temperatures, weakly coupled one another. Both are designed completely alike, but kept in steady states by separate external forcings (periodic vertical acceleration). In such granular gas heat baths, the stationary random motion of the beads resulting from external power supply mimics thermal agitation, characterized by an effective temperature. The mean kinetic energy of the beads is sometimes called 'granular temperature'~\cite{goldhirsh2008}. It is our working hypothesis that the effective temperatures, measured by different means, play the same role as equilibrium temperatures \cite{naert2012,chastaing2017}, in a sense to be discussed below. \\
Let us briefly focus on the granular gas itself. (For reviews, see for instance \cite{goldhirsch2003,brilliantov2004}.) 
The random motion of the inelastic beads is the result of a complex process, in which mechanical power is injected at the bottom and dissipated into heat by collisions and viscous drag. The 'effective temperature' differs from the 'equilibrium temperature' in the sense that the former takes into account only a very reduced number of degrees of freedom, at macroscopic scale, averaging out small scales degrees of freedom. As a result, the values of the effective temperature are definitely distinct. The energy per time unit needed to sustain motion is ultimately completely dissipated into heat. It is the so-called 'housekeeping heat' \cite{oono1998}, for the granular gas. However, the system of interest here is not the  granular gas, but a rotor immersed into it, the former being merely used as a heat bath, causing random forcing on the rotor. \\ 
This model experiment is implemented to investigate specifically the fluctuations of the energy flux $\phi(t)$ between two NESS thermostats kept at distinct effective temperatures $kT_i$ ($i=1; 2$). Working on a macroscopic scale set-up allows to measure conveniently $\phi(t)$ between the baths, and the temperature in each at the same time. \\ 
First, we have a linear dependance of  the mean heat flux in the temperature difference: $\overline{\phi}\propto kT_1-kT_2$. That is the Fourier law for heat conduction~\cite{lecomte2014}. 
Second, the statistics of $\phi(t)$ are examined in terms of the Fluctuation Theorem. Indeed, as the flux is an irreversible transport process between the baths when $kT_1 \neq kT_2$, it causes an asymmetry that can be regarded in these terms. \\
Let us briefly introduce the Fluctuation Theorem (FT), which is a 
cornerstone of the so-called Stochastic Thermodynamics. The FT refers to a set of theoretical results obtained in several steps along the $1990$'s \cite{evans1993, gallavotti1995, kurchan1998, lebowitz1999}. It compares the probability of seeing the entropy of a dynamical or stochastic system increase or decrease, when forced off equilibrium by an external perturbation, with respect to a heat reservoir. It expresses in a very primal expression the irreversibility of a dynamical process:
\begin{equation}
\frac{P\left(\sigma_{\tau}\right)}{P\left(-\sigma_{\tau}\right)}= exp \left(\sigma_{\tau}\right), ~~~~{\rm for~}\tau\rightarrow\infty.
\label{eq0}
\end{equation} 
$P$ is the probability of $\sigma_\tau$, the entropy rate $\sigma=dS/dt$ averaged over a (large) time window $\tau$. The FT is often referred as the most general expression of the $2^{\rm nd}$ principle. Detailed reviews on these important theoretical advances can be found in \cite{seifert2012,jarzynski2011}. Experimental access to observables related to entropy  fluctuations often requires  systems at $\mu$m scale, for molecular thermal fluctuations not to average out. Due to technical difficulties, experimental contributions mostly appeared later. (See a review in \cite{ciliberto2017,ciliberto1999}.) We note however, that some early experiments were performed at macroscopic scale \cite{ciliberto1998}.\\
The entropy being in general not measurable, the relation  $dS=\beta dE$ is used to express the FT for observables such as energy, the bath's inverse temperature is $\beta=\frac{1}{kT}$. \\
It is worth noting that, although the FT holds for system as far as desired from equilibrium, the heat reservoir mentioned above is always implicitly at equilibrium and in the thermodynamic limit, as traditionally assumed in the canonical formalism of the statistical mechanics \cite{kubo1985}. \\
Now, consider that the system, instead of being subject to a deterministic forcing, is perturbed by the coupling with another thermostat. The  result is then a net transfer of energy from one thermostat to the other, and the FT should still apply. 
Indeed, one intuitively expects the back and forth fluxes (negative and positive realizations of $\phi$) to obey the FT. 
Indeed, each system $i$ in contact with a heat bath at $kT_i$ exchanges energy with the other one at a different temperature. We measure $\phi(t)$, the instant resulting energy flux from one to the other and reversely. The time coarse-grained flux is $\phi_{\tau}(t)=\frac{1}{\tau}\int_{t-\frac{\tau}{2}}^{t+\frac{\tau}{2}}\phi(t-t')dt'$. The FT is then, for large $\tau$:
\begin{align}
\frac{P\left(\phi_{\tau}\right)}{P\left(-\phi_{\tau}\right)}= exp\left(\mu \tau \phi_{\tau}\right).
\label{eq1}
\end{align}
The exponent $\mu$ is the only free parameter. 
Jarzynski and W\'ojcik proposed in $2004$ the {\it eXchange Fluctuation Theorem} (XFT), for the heat flux between two equilibrium heat baths \cite{jarzynski2004}. 
According to these authors, the prefactor $\mu$ in Eq.\ref{eq1}, is nothing but the inverse temperatures difference: $\mu=\beta_1-\beta_2=\frac{1}{kT_1}-\frac{1}{kT_2}$.\\
The heat transport has been investigated in terms of the XFT in the past, numerically and experimentally for different kinds of coupling \cite{gomez-marin2006,ciliberto2013,berut2016}, and theoretically for non-Gaussian baths \cite{kanazawa2013}. \\
In a  previous experimental study, we made use of a similar granular gases experiment to investigate the transport between NESS heat baths, weakly coupled by electro-mechanical devices. The Eq.\ref{eq1} was shown to hold precisely. However a quantitative departure from the XFD was observed, as the exponent $\mu$ was significantly distinct from $\Delta\beta$ \cite{lecomte2014}. No explanation was provided at the time. One could invoke the out-of-equilibrium character of the heat baths themeselves, but the dissipative nature of the coupling could also be at play {(an Ohmic resistance in the electric circuit).} \\ 
In the present study, we investigate experimentally the bias due to the dissipative coupling. We show that the XFT is recovered quantitatively in the non-dissipative limit, i.e. ${\mu}\rightarrow{\Delta\beta}$ for vanishing resistance. We stress on the specific feature that the heat baths are in NESS, and not in equilibrium states as implicitly assumed in \cite{jarzynski2004}. \\
The experimental set-up is presented in the next section. 
In Sect.\ref{sect3}, the measurement principle  of the energy flux and temperatures is explained. In Sect.\ref{sect4}, the test of the XFT is presented in the non-dissipative coupling limit, before a short discussion of these results, in Sect.\ref{sect5}.

\section{\label{sect2}
Experiment}
The experimental set-up is composed of two identical granular gas systems. They play the role of two heat reservoirs, coupled one another. Each consists in about $400$ stainless steel beads ($3\,$mm in diameter, $0.1\,$mg mass), steadily shaken vertically at a few g and frequency $40\;$Hz, by an electromagnetic shaker. The beads are contained in $5\,$cm diameter cylindrical cells (see Fig.\ref{fig1}). The probes are $2\,{\rm cm} \times 2\,{\rm cm} \times 0.2\,{\rm mm}$ steel blades, fixed on the vertical shafts of DC micromotors (Maxon, RE 10 118 386). 

\begin{figure}[!h]
\includegraphics[width=55mm]{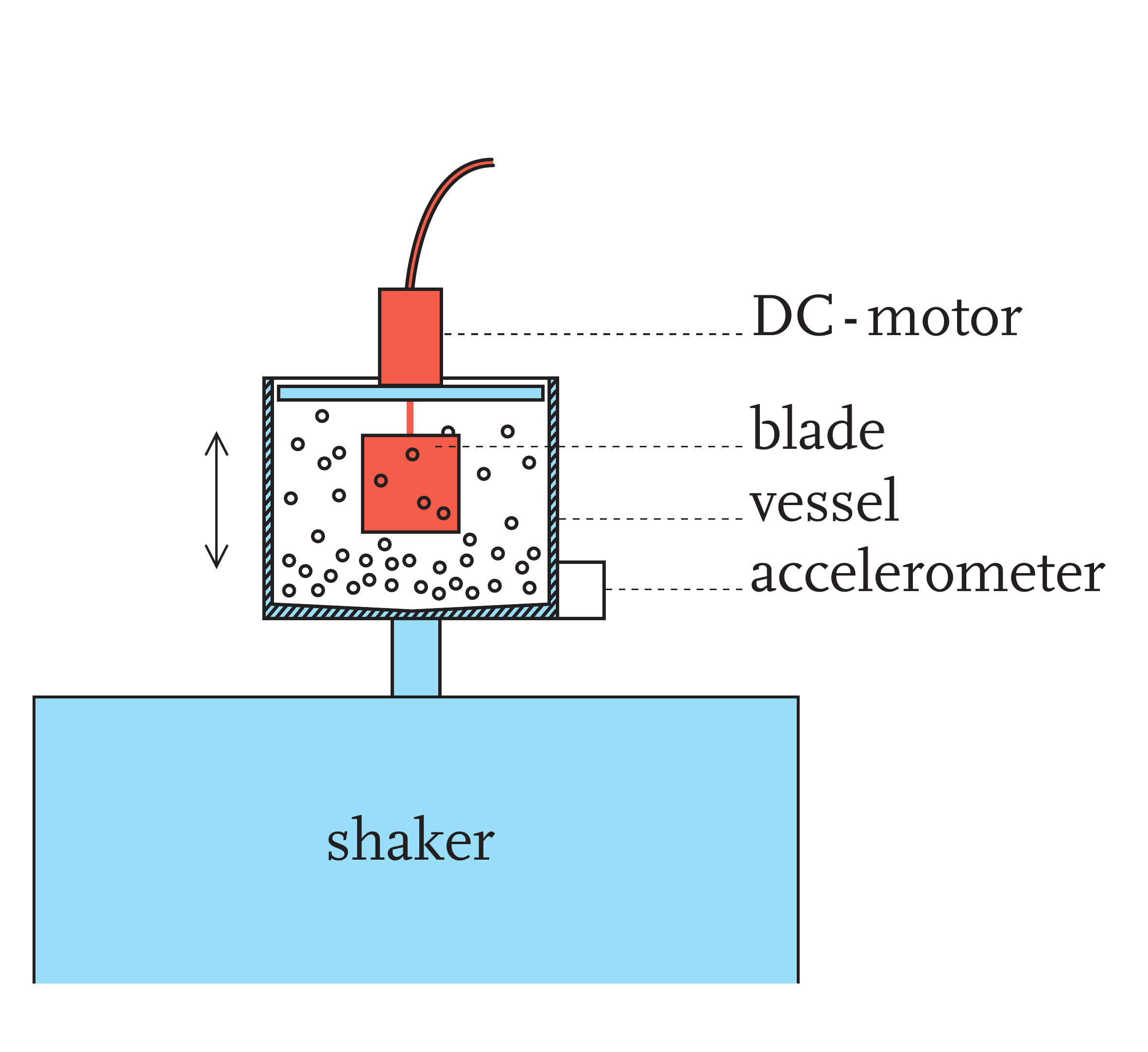}
\caption{\label{fig1} The granular gas is excited in a cylindrical vessel by the vertical acceleration from an electromagnetic shaker. A small blade is fixed on the vertical shaft of a DC micro-motor, set on the immobile cover of the cell. 
Its rotation is caused by the random collisions with the gas.
}
\label{fig1}
\end{figure}

In this configuration, which respects the axial symmetry, the collisions with the gas particles make the blade rotate randomly like a 1D Brownian object \cite{naert2012,monsel2020}. 
A key feature is that a DC motor can also operate symmetrically, as a generator (dynamo). The voltage induced at the terminals, called Electro-Motive Force (EMF), is proportional to the angular velocity: $e(t)=\alpha \dot \theta(t)$ (Faraday's law of induction). The constant~$\alpha \simeq 4.27 \times 10^{-3}\,{\rm}\,$V$\,$s/rad is a characteristic of the device. Conversely, it turns a current $I$ into a torque: $\Gamma(t)=\alpha I(t)$ (Lorentz force), and therefore exchange work with the gas. Note that the constant $\alpha$ is the same in both uses. 
Thanks to this dual function, the DC motor can impose a torque on the rotor imbedded in the gas and measure the velocity, at the same time. It is the only probes used here, that perform measurements on the granular gas. \\
The two probing motors are connected one another by a resistor $R$ (see electrical diagram in Fig.\ref{fig2}). 
\begin{figure}[!h]
\includegraphics[width=78mm]{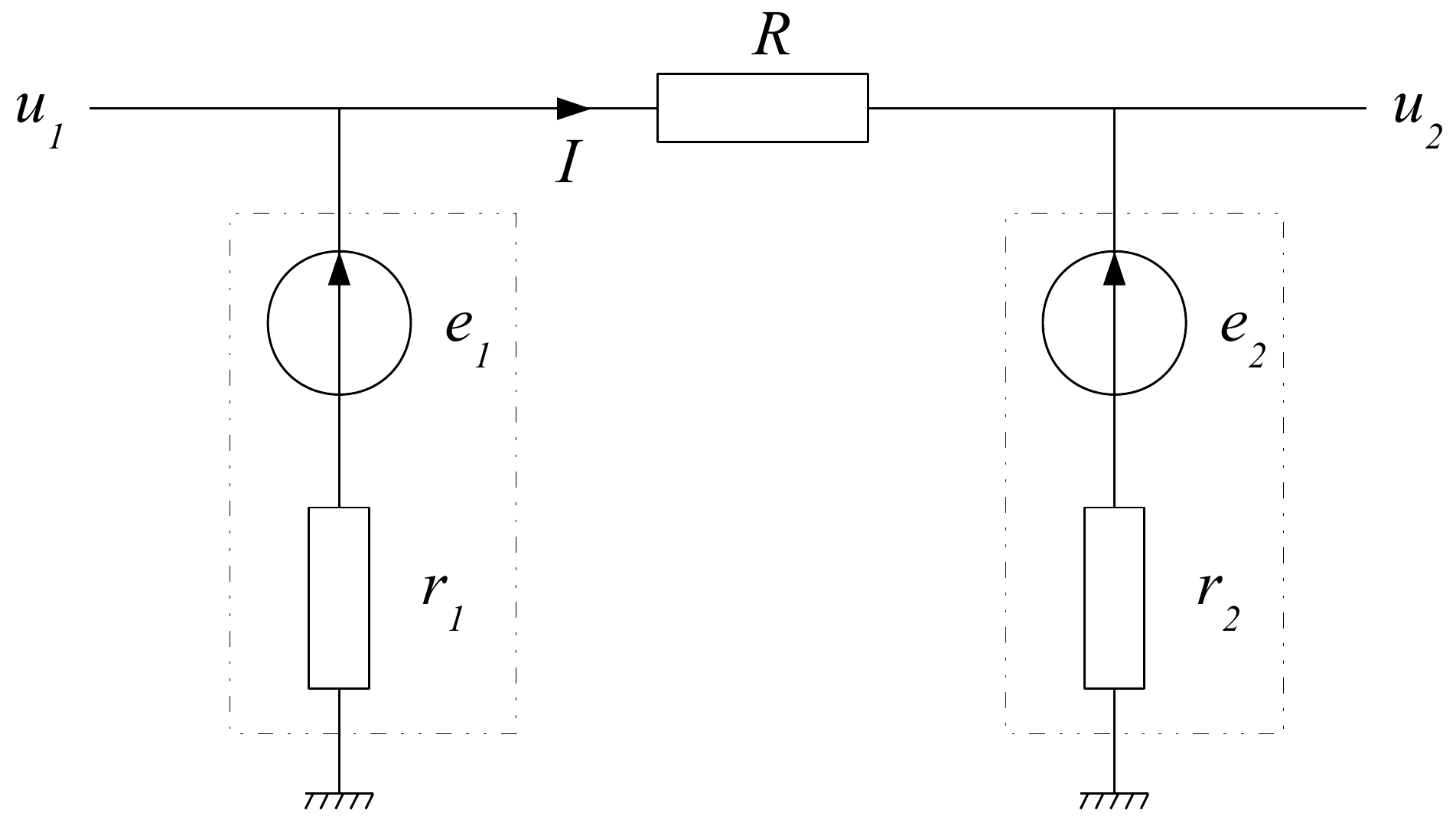}
\caption{\label{fig2} The two motors, coupled to baths $1$ and $2$, are represented in dashed rectangles by ideal voltage sources of EMF $e_1$ and $e_2$ (which instantaneous values are linked to the velocity of rotors), and their internal resistances $r_1$ and $r_2$. 
The coupling resistor is $R$.
}
\label{fig2}
\end{figure}

\noindent The two NESS heat baths are therefore weakly coupled one another. The voltages $u_1(t)$ and $u_2(t)$ are recorded at the terminals of $R$, by a NI-PXI $24$ bits synchronous A/D converter. \\ 
Note that no external power is supplied to the motors: the collisions of the beads on blade $1$ causing its rotation induces a voltage $e_1$. This voltage generates a current $I$ through $R$ into the motor $2$, causing rotation of blade $2$. As a net result, some momentum is transferred from bath $1$ to bath $2$, and reversely. \\
We therefore have at hand an original set-up to investigate the heat transport between two heat baths at different temperatures. One major specificity is that heat baths are in NESS instead of equilibrium states. In the last decade, several contributions has been devoted to verify that the analogy holds, at least as far as stochastic thermodynamics features are concerned. Moreover, the design of the electro-mechanical coupling allows simple and reliable measurements of energy flux as well as temperatures.

\section{\label{sect3}
Heat flux and temperatures measurement
}
The voltages induced in each motor separately, are simply:
\begin{equation} 
\begin{split} 
e_1&=u_1+r_1I, \\ 
e_2&=u_2-r_2I.
\end{split} 
\end{equation}
The internal resistances of the motors are $r_1$ and $r_2$, and the inductances are negligible. The current, constant over the loop, is: 
\begin{equation}
I = \frac{1}{R_{\rm{tot}}}\left(e_1-e_2\right) = \frac{1}{R}\left(u_1-u_2\right).
\label{eq1111} 
\end{equation}
The total resistance of the circuit is $R_{\rm{tot}}=R+r_1+r_2$. The energy flux between $1$ and $2$ is the difference between back and forth fluxes:
\begin{align}
\begin{split} 
\phi(t) & \equiv \left(\dot \theta_1(t) -\dot \theta_2(t)\right) \Gamma(t) \\
&= \left( e_1(t)-e_2(t) \right) I(t).
\end{split}
\label{eq3}
\end{align}
{Noting that at any given moment, each device is virtually either generator or motor, but never both at the same time, conservation of energy implies that the flow is:}
\begin{equation}
\phi(t) = \frac{1}{R_\mathrm{tot}} \left( e_1^2(t)-e_2^2(t) \right).
\label{eq31}
\end{equation}
Besides, the voltages $u_1$ and $u_2$ are sufficient to express temperatures in both baths.  
Two approaches has been proposed to define and measure the effective temperature $kT$. Both rely on  comparing the torque performed (current $I(t)$) and the velocity (EMF $e(t)$). One is making use of the Fluctuation Theorem (FT), the other of the Fluctuation Dissipation Theorem (FDT), both theorems being invoked in a heuristic way. 
The agreement is better that $10\%$. The temperature $kT$ is linked to the kinetic energy of the rotor~\cite{chastaing2017}:
\begin{align}
kT_i &=a+\frac{1}{2} b J\overline{{\dot \theta_i}^2}.
\label{eq111}
\end{align}
$J$ is the moment of inertia ($J\simeq3.33\times10^{-8}\,$kg m$^2$). The constants $a\simeq4\times10^{-7}\,$J and $b\simeq2.5$ must be measured. The relation between the kinetic energy of a 1D free rotor in equilibrium with a thermostat, itself at equilibrium at temperature $kT$, would be $\frac{1}{2}J\overline{\dot{\theta}^2}=\frac{1}{2}kT$. Obviously, such equipartition cannot be assumed in a NESS thermostat. The value of the slope $b\simeq2.5\,(\gtrsim 2)$ reflects the dissipative nature of the granular gas, and the collisions between the beads and the blade. The constant $a$ is the temperature the gas must reach for the beads to get to  the probe. The value $a\neq0$ reflects the fact that the gas is stratified. The coefficients $a$ and $b$ depends on the properties of the granular gas itself, which is not the purpose of the present study. 
Finally, the temperature difference is simply, in electrical variables: 
\begin{equation}
kT_1 - kT_2 = \frac{bJ}{2\alpha^2}
\Bigl(\overline{e_1^2}-\overline{e_2^2}\Bigr).
\label{eq2}
\end{equation}
\noindent The Fourier law for heat conduction is recovered:
\begin{align}
\overline{\phi}&= \frac{1}{R_\mathrm{tot}} \left(\overline{e_1^2}-\overline{e_2^2} \right) \\ 
&= \frac{2\alpha^2}{bJR_\mathrm{tot}} \left( kT_1-kT_2 \right).
\label{eq4}
\end{align}
Note that this expression of the linear response is not exact, in the sense that the coefficient $b$, reflecting the phenomenology of the granular gas, is to be measured.\\
\vspace{.5cm}
\hrule
\vspace{.5cm}

\begin{figure}[!h]
\includegraphics[width=90mm]{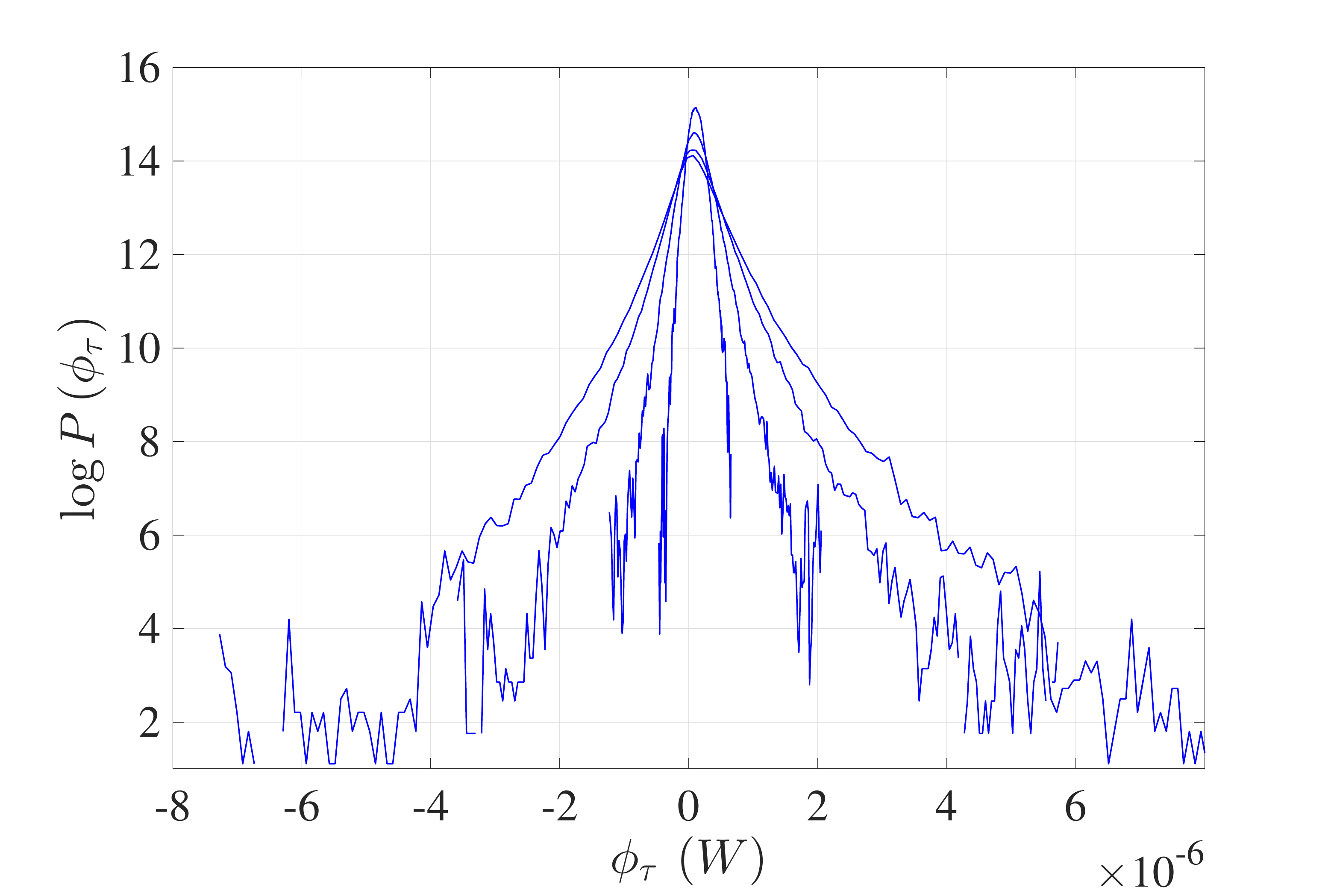}
\caption{\label{fig3} Histograms of the coarse grained heat flux $\phi_\tau$ for several values of $\tau$. (From the widest to the narrowest histograms: $\tau\simeq50\,$ms, $100\,$ms, $300\,$ms, $1.2\,$s.) Here, the average flux is $\overline{\phi}\simeq113\,$nW, for 
 $\Delta\beta\simeq9.37\times10^{4}\,$J$^{-1}$.
}
\end{figure}

\vspace{.2cm}
\indent Let us, from now on, focus on the instantaneous flux $\phi(t)$. It is directly calculated from the  EMF $e_i(t)$ obtained from voltages at the terminals of $R$, as expressed by Eq.\ref{eq31}. 
This flux represents the difference between the heat flux given by bath $1$ and that given by bath $2$. \\
An example of histogram, in Fig.\ref{fig3}, shows that the fluctuations of $\phi$ are indeed asymmetric and intermittent. The time coarse grained $\phi_\tau (t)$ tends to be Gaussian for increasing $\tau$, thanks to the Central Limit Theorem.
{Note that the magnitude of the fluctuations can be of the same order but also much larger than the average energy flow, because, among the huge number of degrees of freedom of such macroscopic systems, only a very few are involved in this transport process.}  Note also that the mean flux can be of the order of $10^{-7}\,$W, or even much smaller. In any case, it is very very little if compared to the power injected into the granular gas by the shakers ($\sim 10\,$W), attesting that the coupling is indeed very weak.

\section{\label{sect4} Test of the XFT in the non-dissipative limit.}
\begin{figure}[!h]
\includegraphics[width=85mm]{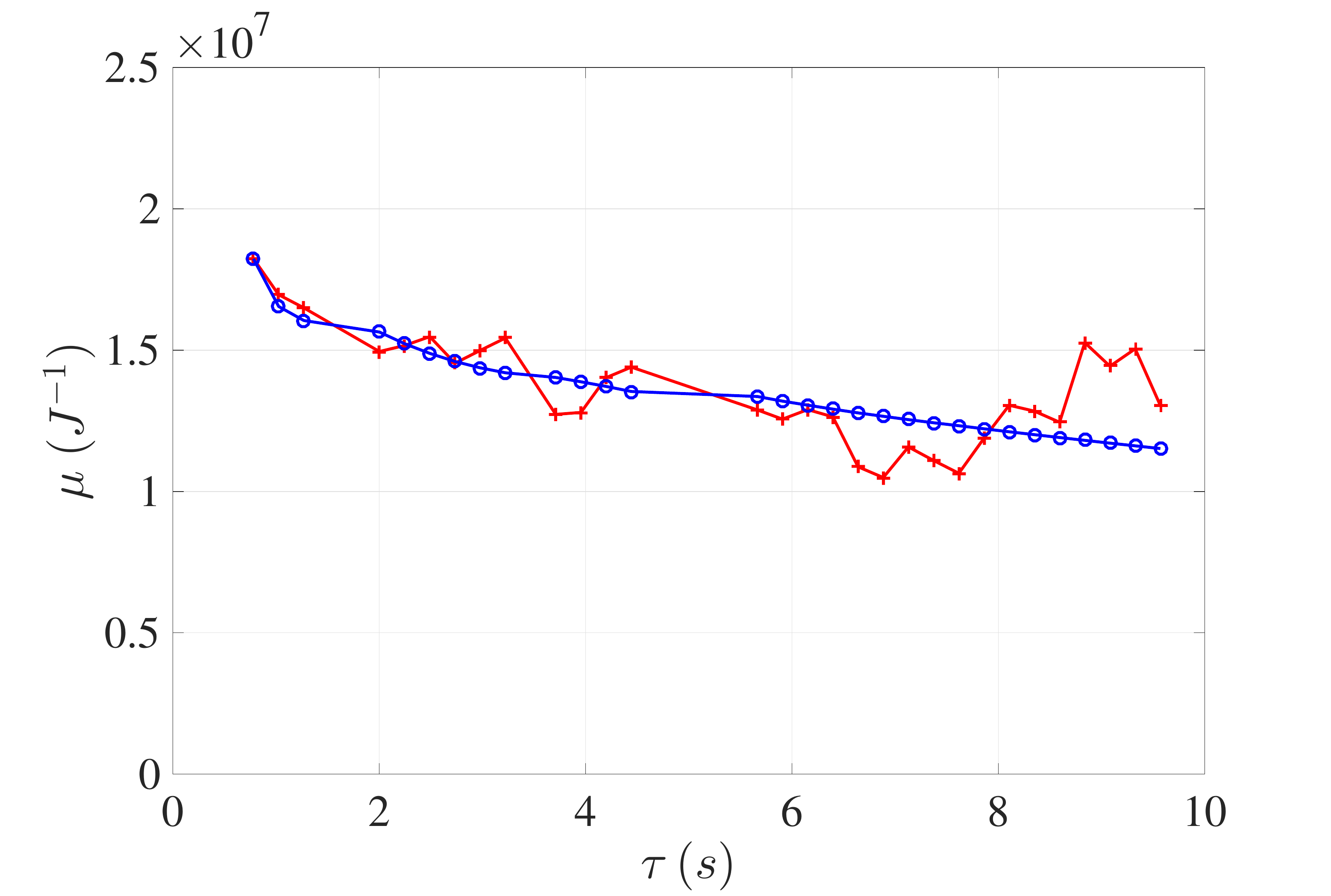}
\caption{
\label{fig4bis}
The slope of the asymmetry function for a $1$h-time sample, obtained from Eq.\ref{eq13}, is plotted  for various times $\tau$ (\textcolor{blue}{$\circ$}), during relaxation. The red curve (\textcolor{red}{$+$}) shows the analysis of the same sample by the best linear fit of the asymmetry function. 
{Here $R=100.23\,\Omega$, and $\Delta\beta\simeq9.37\times10^{4}\,$J$^{-1}$.
}
}
\end{figure}

We have already shown that the heat flux satisfies the equation \ref {eq1}, with a set-up similar to this one \cite{lecomte2014}. {However, the baths being dissipative, the hypothesis for the {XFT} proposed by Jarzynski et al. are not fulfilled. Indeed, the slope $\mu$ of Eq.\ref{eq1} was found quantitatively distinct from $\Delta\beta$, although of the same order.} 
We explore here the following idea: vary the resistance $R$ and compare the subsequent values of $\mu$. Extrapolating $\frac{\mu}{\Delta\beta}$ versus $R_{\rm{tot}}$ gives the 'non-dissipative coupling limit', when $R_{\rm{tot}} \rightarrow 0$. \\
Previously, $\mu$ was measured thanks to the slope of the asymmetry function $\delta(\phi_{\tau})~=~\frac{1}{\tau}\log\frac{P(\phi_{\tau})}{P(-\phi_{\tau})}$ { versus} $\phi_{\tau}$. The asymptotic value of the slope $\mu$, in the large time limit $\tau \rightarrow \infty$, were compared to $\Delta\beta$. This protocol is difficult to implement in a stable manner. For instance, it is sensitive to the large range of conditions we address here (various $R$ and $\Delta kT$), and to the sample size. In the end, the uncertainties are difficult to control. \\
Here we have used another method to measure the same quantity. This method was found fruitful in another context \cite{apffel2019}, in the sense that it is more practical. It is strictly equivalent in the limit of large $\tau$. It is based on the ratio of two first moments of the flux: 
\begin{equation} 
\begin{split} 
\mu=\frac{\tau}{2}\frac{\overline{\Delta\phi_{\tau}^2}}{\overline{\phi_{\tau}}}, ~~~\rm{for~~} \tau \rightarrow \infty.
\end{split} 
\label{eq13}
\end{equation}
$\overline{\phi_{\tau}}$ is the mean and $\overline{\Delta\phi_{\tau}^2}$ is the variance of the flux. This alternative method, efficient and obviously much easier in practice, is valid even for variables like $\phi_{\tau}$ which are Gaussian only for the largest $\tau$. Convergence is easier because it involves low order moments only, which counts at large $\tau$ (see Fig.\ref{fig4bis}). Also, the ratio \ref{eq13} can be calculated even though negative values of $\phi_\tau$ are absent.

\begin{figure}[!h]
\includegraphics[width=90mm]{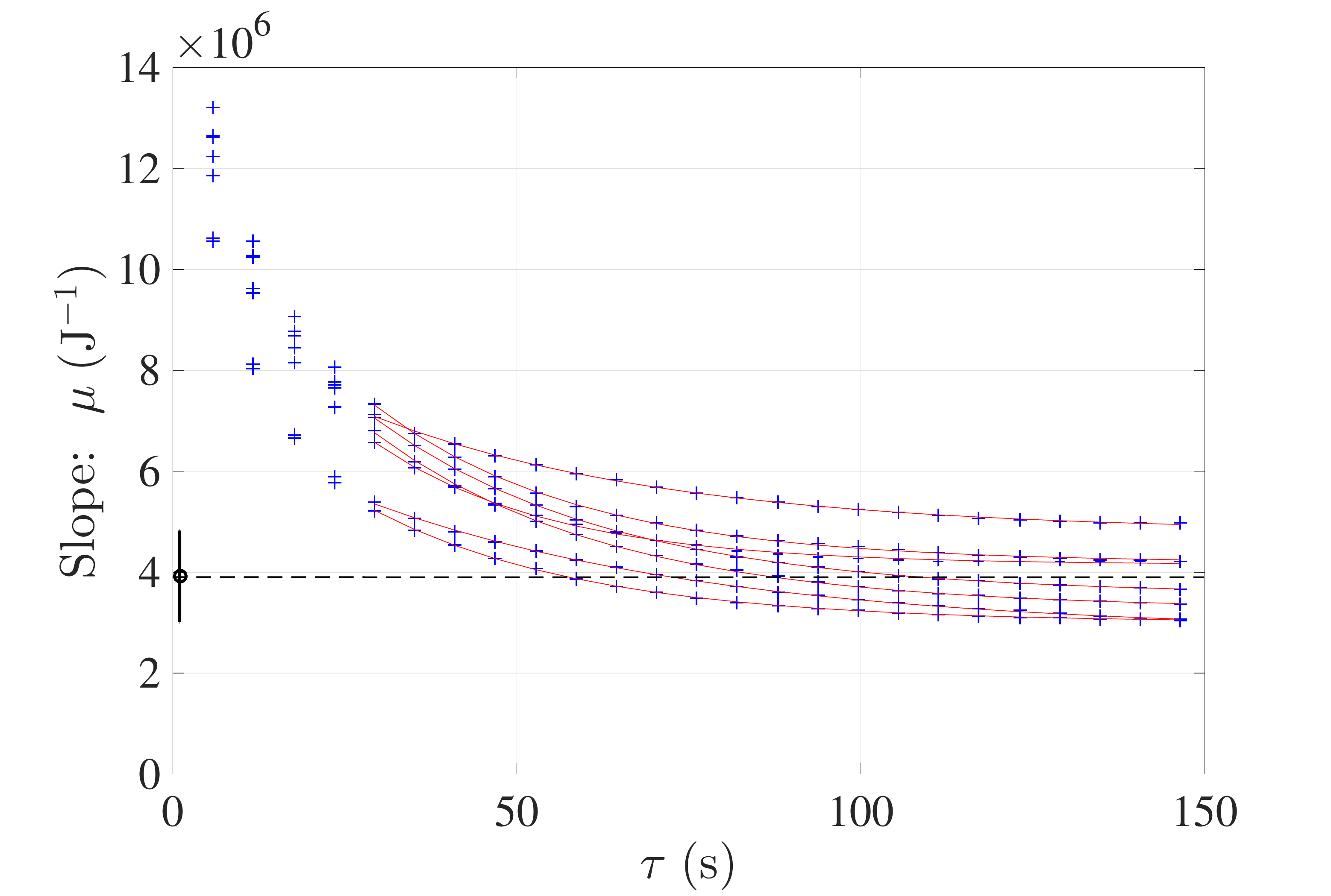}
\caption{\label{fig4} The slope of the asymmetry function for a few time series, calculated  thanks to Eq.\ref{eq13}, is plotted  for various time-lags $\tau$ (\textcolor{blue}{$+$}). The red curves show best exponential fits, 
The dashed black line represents the average of these limits at large $\tau$ over dozens of time series. RMS of $\mu$ is in black. %
Here $R=23.28\,\Omega$, while $\Delta\beta\simeq1.5\times10^5\,$J$^{-1}$.
}
\end{figure}

It is known that the relaxation at finite time is not universal \cite{joubaud2007,ciliberto2010}. 
Now, as we are only interested in the $\tau\rightarrow\infty$ asymptotic value of $\mu$, what  is the most precise and reliable way to evaluate it? A robust protocol must be defined, unique to all conditions of interest  here: all temperature gradients $\Delta kT$, at various $R$. In the large $\tau$ limit, we found convenient to use an exponential fitting to obtain the value of the asymptote in a stable and reproductible way. The ratio given is Eq.\ref{eq13}, $\frac{\tau}{2}\frac{\overline{\Delta\phi_{\tau}^2}}{\overline{\phi_{\tau}}}$,  
is calculated for various times $\tau$, for a large number of samples and plotted versus $\tau$ in Fig.\ref{fig4}, together with an exponential best fit: $\mu+c\,exp(-\tau/\tau^*)$. 

\begin{figure}[!h]
\includegraphics[width=85mm]{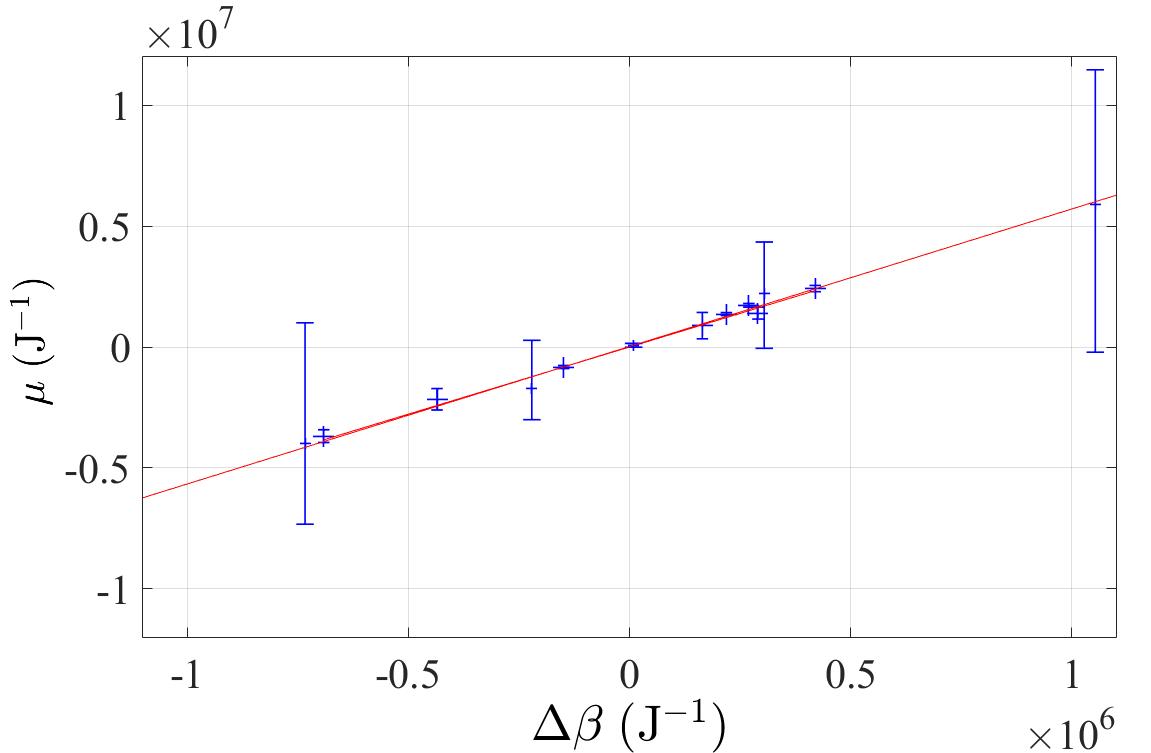}
\caption{\label{fig5}This figure is taken from Ref.\cite{lecomte2014}: the relation between $\mu$ and $\Delta\beta$ is linear, for $R=22\,\Omega$, and the slope is $\frac{\mu}{\Delta\beta}\simeq 5.69$. }
\end{figure}

The agreement, although not perfect, is acceptable for any configuration of interest here. Averaging the limit value obtained from this fitting over dozens of time series gives a measure of the slope $\mu$. The RMS gives an estimate of the uncertainties. \\
Effective temperatures $kT_i$ in both thermostats are calculated thanks to the variances of the EMF, $e_i(t)$, and the Eq.\ref{eq111}, that is to say, using the coefficient $b$ obtained in \cite{chastaing2017}. As mentioned in Lecomte et al., the relation between $\mu$ and $\Delta\beta$ is linear for the value $R=22\,\Omega$. For self-consistency of the paper, the main figure from Lecomte et al. is reproduced in Fig.\ref{fig5}. It was noted at the time that the slope was not $1$, in contrast to the XFT. 

\noindent Acknowledging this proportionality, we performed new measurements for different resistances $R$, only for a few values of temperature difference to ascertain the variation of the slope. \\

\begin{figure}[ht]
\includegraphics[width=85mm]{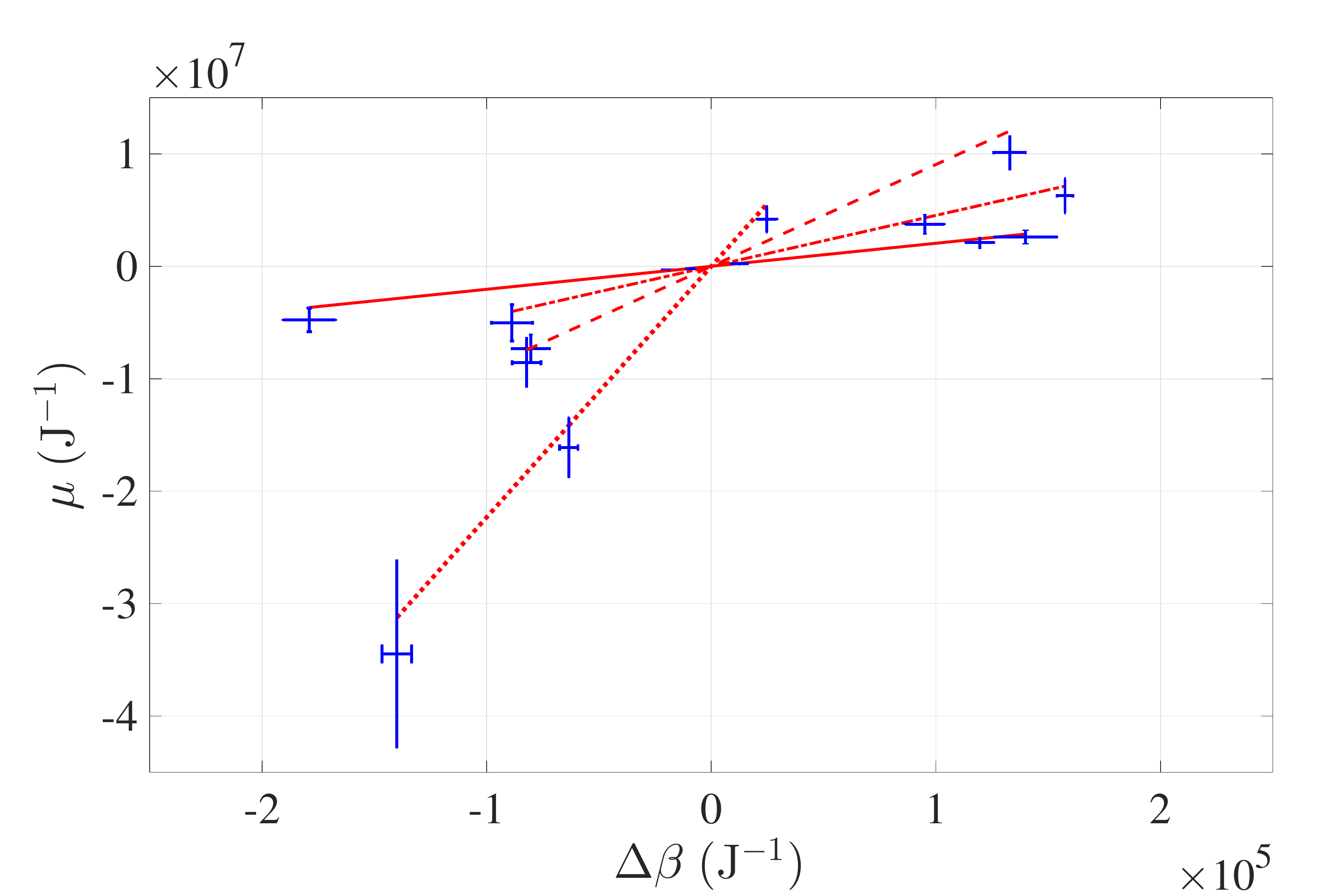}
\caption{The slope $\mu$ is plotted against $\Delta\beta$ for a few values of the coupling resistance $R=23.27\,\Omega\,$(solid)$;\,100.23\,\Omega\,$(dash-dot)$;\,242.6\,\Omega\,$(dash);$\,617.4\,\Omega\,$(dot). The red lines represent in each case the best linear fits through $0$. 
\label{fig6}
}
\end{figure}

Several measurements of $\mu$ are plotted against $\Delta\beta$, superimposed in Fig.\ref{fig6}. These results show that the proportionality coefficient between $\mu$ and $\Delta\beta$ decreases monotonously with $R$.\\ 
\noindent To be fair, one must consider the total resistance in the loop, to account for the total dissipation in the coupling, $R_{\rm{tot}}=R+r_1+r_2$. The dependance of $\frac{\mu}{\Delta\beta}$ in $R_{\rm{tot}}$, is presented in Fig.\ref{fig7}. 
In order to extrapolate to $R_{\rm{tot}}=0$, a polynomial fitting is performed ($2^{\rm nd}$ order is sufficient). It leads to the slope $\frac{\mu}{\Delta\beta}\simeq 0.85$, in the non-dissipative limit. This value is compatible with unity, as expected by the XFT. Here is the main result of this study.\\

\begin{figure}[!h]
\includegraphics[width=85mm]{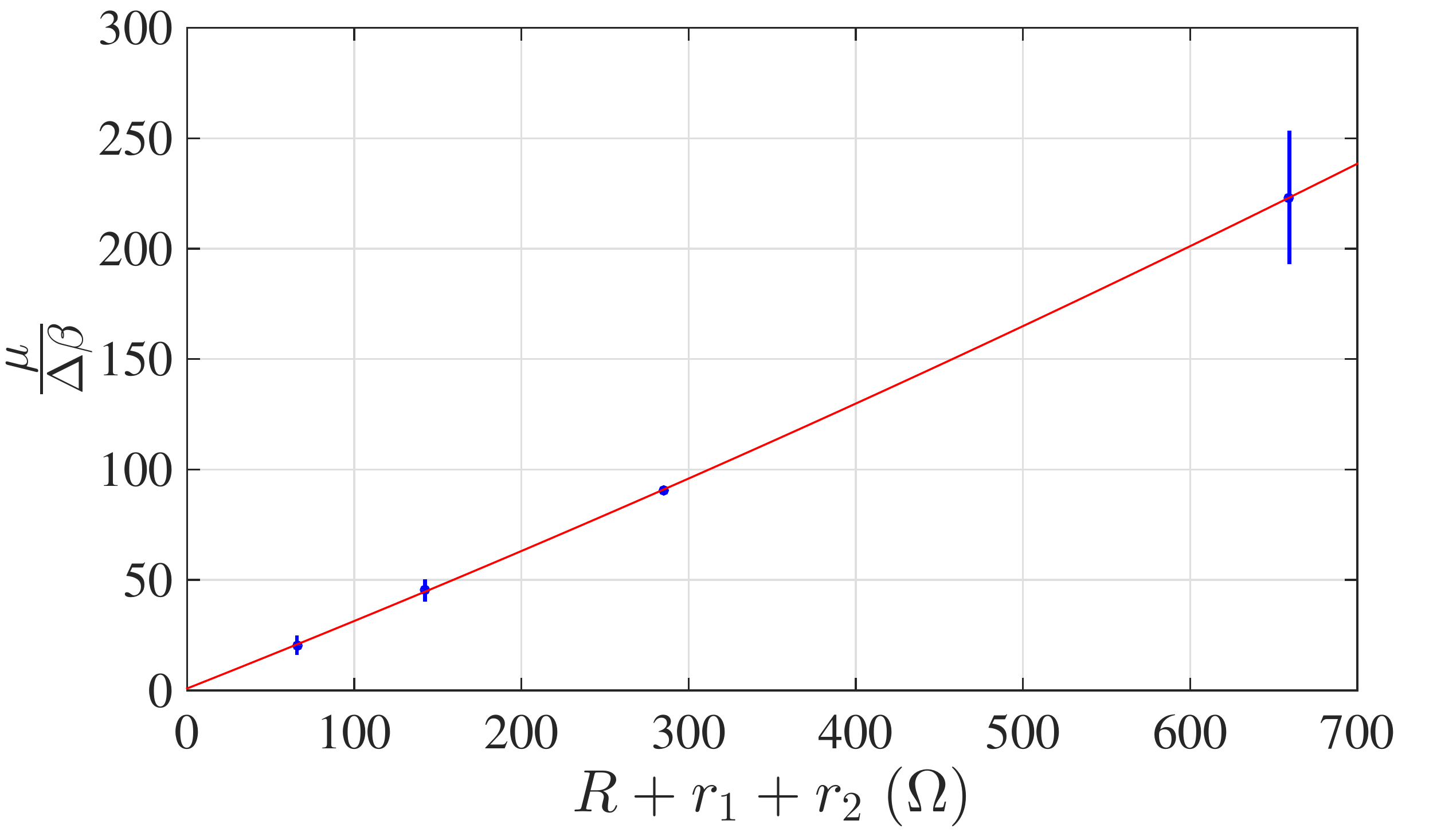}
\caption{\label{fig7}The ratio $\frac{\mu}{\Delta\beta}$ is plotted against the resistance $R_\mathrm{tot}$. 
A 2\textsuperscript{nd} order polynomial fitting is performed $y=p_1\,x^2 + p_2\,x + p_3$ (in red). The coefficients obtained are $p_1\simeq5.63\times10^{-5};\,p_2\simeq0.30;\,p_3\simeq0.85$. In the zero-resistance limit $\frac{\gamma}{\Delta\beta}\rightarrow~p_3\simeq0.85$.
}
\label{fig7}
\end{figure}

\indent At this point, a discussion of the uncertainties is pivotal. Some of the errors on $\mu$ are due to the exponential fitting process presented above to determine the asymptote, and statistic limitations. An additional source of error comes on $\Delta\beta$ from the limited  accuracy of the coefficient $b$, barely  better than $10 \%$. Minor errors come from the drift in room temperature, like night/day variations (inducing variations of $\sim1\%$ at most on $r_i$). The error bars on $R_\mathrm{tot}$ are imperceptible, see Fig.~\ref{fig7}. Note that the increase of the error bars with $R$ is due to the currents decrease, deteriorating the S/N ratio. All this leads to an uncertainty on the ratio $\frac{\mu}{\Delta\beta}$, overall estimated to about $1$ at most. In the end, it is a fraction of a percent in relative value, which is satisfactory.

\section{\label{sect5} Discussion/Conclusion.}
We have presented a continuation, until concluding, of the preceding work of Lecomte et al. in $ 2014 $ on the transport of heat between two NESS thermostats maintained at different effective temperatures $kT_i$. Two identical cm-size 1D-Brownian mobiles, fastened on electric micro-motors, are immersed in granular gas heat baths. The motors are electrically linked one another by a resistance $R$, insuring a weak coupling between the baths. Electrical measurements at the terminals of $R$ give access to the flux $\phi(t)$ and the temperature in each bath, $kT_i$. In the present study, the coupling is varied by changing $R$. For each value, the ratio of the variance over the mean flux are calculated. This is equivalent to the slope of the asymmetry function, for several temperature differences. We show experimentally that, extrapolating to the non-dissipative coupling limit ($R \rightarrow 0$), the fluctuations of the flux are compatible with the eXchange Fluctuation Theorem (XFT) proposed by Jarzynski et al. in $2004$, in the large time limit: 
\begin{align}
\frac{P\left(\phi_{\tau}\right)}{P\left(-\phi_{\tau}\right)}= exp\left(\Delta\beta \tau \phi_{\tau}\right),
\label{eq11}
\end{align}
$\phi_{\tau}$ being the time coarse grained heat flux. The inverse temperature difference is $\Delta\beta=\frac{1}{kT_1}-\frac{1}{kT_2}$. \\\\
\indent Whereas the Fourier law for heat transfer links the mean flux and temperature gradient, the XFT expresses the asymmetry of the fluctuations, amongst non-equilibrium as well as equilibrium finite size subsets. \\
{There is several interesting aspects to the experimental observing of the XFT in such a context, where it is unexpected. 
A first question that requires careful reflexion is where non-equilibrium actually stands. The XFT, like other FT, applies to such out-of-equilibrium situation where heat flows between two systems. It is intrinsically a non-equilibrium feature. Consequently, having the flux between 2 baths coupled verifying XFT as well as Fourier law looks finally unsurprising. The concern is more acute about the dissipative nature of the granular gas itself. (See a questioning of the FT in this context in \cite{visco2005}). The NESS of the granular gas results from the balance between work injection and heat dissipation. Once in a steady state of fluctuations, the Brownian rotor experiences kicks, but has no way to make any difference to whether this random forcing results from an equilibrium state or merely a NESS. Mathematically, the coupling with the bath is represented by a random forcing term in the equation of motion of the rotor. It has no signature of the irreversibility of the underlying work to heat conversion process.\\ 
Another question concerns the effective temperature $kT$, that characterizes the agitation of the beads. The verifications of the laws of heat transport in a gradient of $kT$, for average as well as fluctuations, confirms once more that $kT$ behaves the same way as an equilibrium temperature. This second observation looks even more striking! But it is just another aspect of the previous one: NESS or equilibrium state heat baths are not the same, indeed: they just behave the same from the point of view of applying stochastic thermodynamics on the rotor. This is obviously not true for the motion of the beads in the granular gas!} \\
Intriguing deductions are starting to emerge from these results. First of all, the field of application of stochastic thermodynamics appears not to be limited to the molecular scale. Some open questions on macroscopic scale phenomena, are not addressed by statistical physics because they are out of equilibrium, that is, dissipative and spatially extended systems, such as those of interest in nonlinear physics. Yet, it seems possible to invoke the theoretical arsenal of stochastic thermodynamics as a wedge in long-lasting problems, such as hydrodynamic turbulence, wave turbulence, rapid fracture in solids, granular matter, etc. 

\begin{acknowledgments}
We are grateful to S.~Auma\^itre, C.~Crauste, A.~Archambault, A.~Steinberger, and J.-P.~Zaygel for many discussions and suggestions in the course of the experiments, and also to J.-C.~G\'eminard, M.~Adda-Bedia and C.~Jarzynski for corrections and suggestions on the manuscript.

\end{acknowledgments}

\appendix

\end{document}